\documentclass[prl,showpacs,12pt]{revtex4}

\usepackage{graphicx}
\usepackage{dcolumn}
\usepackage{amsmath}
\usepackage{epsfig}

\preprint{xxx/xxx-xxx}

\begin{document}

\title{Asymmetrical Quantum Cryptographic Algorithm}
\author{Guihua Zeng$^1$}
\email{guihuazeng@hotmail.com}
\author{Carlos Saavedra$^{1,2}$}
\email{carlos.saavedra@udec.cl}
\author{Christoph H. Keitel$^1$}
\email{keitel@uni-freiburg.de}
\affiliation{ $^1$ Theoretische Quantendynamik, Fakult\"at f\"ur
Physik, Universit\"at Freiburg,
Hermann-Herder-Stra{\ss}e 3, D-79104 Freiburg, Germany \\
$^2$ Departamento de F\'{\i}sica, Universidad de Concepci\'on,
Casilla 160-C, Concepci\'on, Chile}

\date{\today}

\begin{abstract}
A cryptographic algorithm is proposed based on fully quantum
mechanical keys and ciphers. Encryption and decryption are carried out via an
appropriate measurement process on entangled states as governed by
a quantum mechanical, asymmetrical and dynamical public key
distribution. The use of public keys leads to a high availability
of our scheme, while their quantum nature is shown to ensure
unconditional security of the proposed algorithm.
\end{abstract}


\maketitle

\newpage

In quantum cryptography \cite{wiesner83} messages are rendered unintelligible to unauthorized users via quantum
mechanical means, i.e. a quantum algorithm. A cryptographic algorithm in general describes the encryption and decryption
mechanisms while the keys involve all necessary additional information. So far, many aspects in quantum cryptography
have been studied such as quantum key distributions \cite{bennett84,ekert91,lo99, bennett92}, quantum secret sharing
\cite{hillery99}, quantum identity verification \cite{zeng00}, quantum bit commitment and quantum multi-party
computation \cite{kent99}, quantum information hiding \cite{terhal01}, and information theory for quantum cryptography
\cite{schumacher98}. Present-day quantum cryptography involves quantum keys and classical cryptosystems, which are
both well understood and implemented experimentally \cite{schneier94}. The classical cryptosystem can be categorized
as classic symmetrical key cryptosystem (SKC) and classic asymmetrical i.e. public key cryptosystem (PKC) \cite{diffie76}. The
characteristic of the SKC is that encryption and decryption use the same key (called symmetrical key), which are kept
secret by the communicators. The main feature of the PKC is that the public key associated with a private key can be
published. By the public key one can not in principle obtain any information about the private key. Since the holder
may publicly announce the public key, everyone who wants to communicate with the holder can easily find and use it.
Classic cryptographic algorithms have been widely used in both private information protection and private
communication.

There are drawbacks, however, in both classic SKC and PKC.
Currently, the one-time pad is the only algorithm which has been
proven secure, but it can not be used efficiently in practical
applications because of difficulties in the key management. Although
the protocols for quantum key distributions provide an efficient
way, the problem of availability of the one-time pad cryptosystem has
not been completely solved, because the classic SKC can not be used
efficiently in large network systems. The classic PKC, which was
proposed 20 years ago, can provide high availability for the
cryptosystem. However, since the classic PKC relies on the assumption
of computational complexity such as the difficulty of
factoring large numbers, up-to-date none of the existing classic
PKC has been proven secure, even against an attacker with limited
computational power. In additional, the rapid development of
quantum computers \cite{chuang95,shor94,cirac95} increasingly
endanger the security of current cryptosystems. Research shows
for example that a quantum computer may easily break the well-known
RSA algorithm \cite{shor94}.

In this letter we introduce a quantum public key algorithm. The algorithm makes use of maximally entangled states
(MES) of pairs of spin-$\frac{1}{2}$ particles and their correlation-preserving projection on appropriate directions.
It begins with the generation of public and private keys via the correlation among MES, associated measurement operators
and a string of unitary operators. Then the sender, Bob, encrypts his message by using the public key and a quantum logic 
gate operation which is governed by the key and yields the ciphertext. Finally, the private key is employed by
Alice to decrypt this ciphertext. The unconditional security and availability of the proposed algorithm are shown to be
guaranteed, respectively, by the no-cloning theorem \cite{wooters82} and by the technology of the public key.

The central and difficult problem of designing a public key algorithm is how to
generate the secure key pairs, i.e., the public key and the secure private
key. For the quantum key generation and distribution, many protocols have been
proposed. However, in all previous schemes only symmetrical keys can be
generated and distributed, so those previous protocols for quantum key
generation and distribution can only be used in the SKC but are not suitable
for the PKC.

We here present a secure key distribution for our quantum PKC via using maximally entangled states of pairs of
spin-$\frac{1}{2}$ particles. The single-particle eigenstates are denoted $|0\rangle $ and $|1\rangle $ 
with respect to a measurement along an axis $z$, i.e. $\sigma _{z}=|1\rangle \langle 1|-|0\rangle \langle 0|$, 
and $|\pm \rangle = (|0\rangle \pm |1\rangle )/\sqrt{2}$ are the eigenstates of the spin operator along the 
corresponding $x$ axis, i.e. $\sigma _{x}= (|1\rangle \langle 0| +|0\rangle \langle 1|)/2$.
We consider the so called Bell states  $|\Phi ^{\pm }\rangle =(|00\rangle \pm |11\rangle )/\sqrt{2}$ and
$|\Psi ^{\pm }\rangle =(|01\rangle \pm |10\rangle )/\sqrt{2}$ and the additional MES given by 
$|\psi ^{\pm }\rangle =(|0-\rangle \pm |1+\rangle )/\sqrt{2}$ and $|\phi ^{\pm}\rangle =(|0+\rangle \pm |1-\rangle )/\sqrt{2}$. 
We shall refer to all these two-particle MES as quantum channels. These states can be generated by applying unitary 
transformations on one of the particles of any of above MES, keeping the degree of entanglement unchanged. 
It will turn out beneficial to express the considered MES in the various bases offered by the sets of eigenstates 
of spin operators in the various directions, for example 
\begin{eqnarray}
|\Phi ^{+}\rangle  &=&|0,0\rangle +|1,1\rangle =|+,+\rangle +|-,-\rangle ,
\notag \\
|\phi ^{+}\rangle  &=&|0,+\rangle +|1,-\rangle =|+,0\rangle +|-,1\rangle ,
\label{eq:basis}
\end{eqnarray}
where the normalization factors have been omitted. 

From equation (\ref {eq:basis}), it can be noted that, if the spin
of one particle of the MES $|\Phi ^{+}\rangle$  is being measured 
along the axis $x$ or $z$, the state of the other particle is completely 
determined when its spin is also measured along the same axis. This can
be generalized easily to any axis $\hat{n}$. For the state 
$|\phi ^{+}\rangle $, we find however, that the particles must be measured
in orthogonal directions. In table~\ref{tab:correl} it is shown how
this situation is for all considered MES for measurements along the 
$x$ and $z$ axes.
\begin{table}[tbp]
\caption{Measurement axis are indicated for each particle of a
MES, along the columns, for obtaining maximum correlation
 or anti-correlation between the readouts of measurements.}
\label{tab:correl}%
\begin{ruledtabular}
\begin{tabular}{||l||c|c||c|c||}
 Quantum channel  & \multicolumn{2}{c||}{$|\Phi^\pm\rangle,\, |\Psi^\pm\rangle$} &
                    \multicolumn{2}{c||}{$|\phi^\pm\rangle,\, |\psi^\pm\rangle$ }\\
\hline
 Particle 1 ($\mathcal{M}_{P}$) & $\sigma_z$ & $\sigma_x$ & $\sigma_z$ & $\sigma_x$ \\
\hline
 Particle 2 ($\mathcal{M}_{S}$) & $\sigma_z$ & $\sigma_x$ & $\sigma_x$ & $\sigma_z$ \\
\end{tabular}
\end{ruledtabular}
\end{table}
Thus, we learn from table~\ref{tab:correl} that a given quantum 
channel and the measurement axis for both particles are correlated, 
i.e. if two of them (including the channel) are known, the third can 
be determined. However, if only one is known, the other parameters remain unknown. Based 
on this feature we shall continue in constructing the public key 
$K_{P}$ and the corresponding private key $K_{S}$.

Alice initiates the key generation by choosing on paper random strings of 
both quantum channels $\mathcal{B}=\{|b_{1}\rangle ,|b_{2}\rangle ,\cdots
,|b_{n}\rangle \}$ and spin operators for one particle 
$\mathcal{M}_{P}=\{m_{p}^{1},m_{p}^{2},\cdots ,m_{p}^{n}\}$
 with $|b_{i}\rangle \in \left\{ |\Phi ^{\pm }\rangle ,|\Psi ^{\pm
}\rangle ,|\psi ^{\pm }\rangle ,|\phi ^{\pm }\rangle \right\} $
and  $m_{p}^{i}\in \{\sigma _{z},\sigma _{x}\}$
(later the quantum channels will arise from actual experiments).
Following  table~\ref{tab:correl} Alice is now in the position to 
determine the spin measurement axis with regard to the second particle,
yielding $\mathcal{M}_{S}=\{m_{s}^{1},m_{s}^{2},\cdots
,m_{s}^{n}\}\,$. Then, Alice creates an additional string of
unitary operators $\mathcal{U}=\{U_{1},U_{2},\cdots ,U_{n}\}$, where 
$U_{i}=\cos \theta _{i}\left(|0\rangle \langle 0|+|1\rangle \langle 1|\right) 
+\sin \theta _{i}\left(|1\rangle \langle 0|-|0\rangle \langle 1|\right) $ 
with $\theta _{i}$ being a random number, which is secretly chosen by Alice. 
Combining $\mathcal{M}_{P}$ and $\mathcal{U}$, Alice is then able to generate 
our public key $K_{P}$,
\begin{equation}
{K}_{P}=\{k_{p}^{1},k_{p}^{2},\cdots ,k_{p}^{n}\},\qquad
k_{p}^{i}=U_{i}^{-1}m_{p}^{i}U_{i}.  \label{eq:public}
\end{equation}
Thus, a spin measurement operator $k_{p}^{i}=\sigma _{\hat{n}_{i}}$ 
along an axis $\hat{n}$ may be publicly announced while
the quantum channel and the measurement operator on the second particles
of the quantum channels remain known solely to Alice. 
The corresponding private key $K_{S}$ is then constructed via
\begin{equation}
{K}_{S}=
\{k_{s}^{1},k_{s}^{2},\cdots ,k_{s}^{n}\},\qquad
k_{s}^{i}=U_{i}^{-1}m_{s}^{i}U_{i}.
 \label{eq:secret}
\end{equation}
where $k_{s}^{i}=\sigma_{\hat{n}_{i}}$ for Bell states 
$|b_i\rangle$, $k_{s}^{i}=\sigma _{\hat{n}_{i}^{\perp }}$ 
for the other employed MES and $\hat{n}_{i}^{\perp }$ being 
an orthogonal direction to $\hat{n}_{i}$.
This relation among $k_{p}^{i}$ and $k_{s}^{i}$ was 
derived like table~\ref{tab:correl} but with $x$ and $z$ 
replaced by the general directions $\hat{n}_{i}$ and 
$\hat{n}_{i}^{\perp }$.

The secret key ${K}_{S}$ is dependent on the parameters 
$\mathcal{M}_{S}$, $\mathcal{B}$ and $\mathcal{U}$, so that we shall refer 
to it as a dynamical key. This becomes relevant in practical applications 
because it has been proven that dynamical keys are more secure than static keys. 
There is no way that the secret private key ${K}_{S}$ can be determined 
with the mere knowledge of the public key ${K}_{P}$, because both the 
quantum channel and the unitary rotation still remain unknown to everybody but Alice. 
The private key ${K}_{S}$ is kept secret by the holder while the public key 
${K}_{P}$ may be published like a telephone number. The use
of the public key leads to a high availability for the proposed
scheme. At the same time, the high secrecy of $\mathcal{M}_{S}$
and $\mathcal{B}$ leads to a high secrecy for the private key. We note that up 
to this point all procedures may be carried out on paper, while in what comes an actual 
experiment is required.

With regard to the encryption and decryption procedures, Alice and Bob are imagined
to share particles of a set of $m$ identical MES $|\Phi ^{+}\rangle $ with $m>n$ at 
this stage. 
One particle of each MES is associated with Alice and one with Bob which form 
the one-particle strings $\mathcal{P}_{A}^{^{\prime }}$ and 
$\mathcal{P}_{B}^{^{\prime }}$, respectively.
The labels $A$ and $B$ refer to Alice's and Bob's particles throughout the article.
Then Alice and Bob choose respectively a fraction of particles
(denoted by $\Delta \mathcal{P}_{A}^{^{\prime }}$ and $\Delta 
\mathcal{P}_{B}^{^{\prime }}$, respectively) from the sets 
$\mathcal{P}_{A}^{^{\prime }}$ and $\mathcal{P}_{B}^{^{\prime }}$ 
to check on eavesdropping by using the method presented in Ekert's 
protocol for quantum key distributions \cite{ekert91}. 
Whenever eavesdropping has occurred, it is necessary
to establish again the string of quantum channels.
Otherwise, the remaining entangled states may be arranged to have 
$n$ states and form the set $\mathcal{B}^{\prime}$. 
For convenience, we denote the remaing particles
as $\mathcal{P}_{A}=\mathcal{P}_{A}^{^{\prime }}-\Delta \mathcal{P}_{A}^{^{\prime}}=
\{p_{A}^{1},p_{A}^{2},\cdots ,p_{A}^{n}\}$, and $\mathcal{P}_{B}=
\mathcal{P}_{B}^{\prime}-\Delta \mathcal{P}_{B}^{\prime}=\{p_{B}^{1},p_{B}^{2},
\cdots ,p_{B}^{n}\}$.
Then Alice generates a set $U_A= \{U_{A 1}, \cdots , U_{A n}\}$ by randomly choosing
$U_{A i} \in \left\{ I,H,\sigma _{z},H\sigma_{z},\sigma _{x},H\sigma _{x},\sigma _{y},H\sigma _{y}\right\}$
for $i \in \{1, \cdots, n\}$ and thus creates $\mathcal{B}=\{U_{A 1} |\Phi ^{+}\rangle, 
\cdots, U_{A n} |\Phi ^{+}\rangle \}$.  Here $I$ is the identity operator,
$H=\left( |0\rangle \langle 0|+|1\rangle \langle 1|+|1\rangle
\langle 0|-|0\rangle \langle 1|\right) /\sqrt{2}$ is a Hadamard
gate and we have neglected a global phase. 
As an example $|\Psi^{+}\rangle =$ $\sigma_{x_{A}}|\Phi ^{+}\rangle$ 
and $|\phi^{+}\rangle =H_{A}|\Phi ^{+}\rangle $  where the subindex A 
indicates that the corresponding operator need be applied on Alice's particle. 
Then Alice has obtained the set $\mathcal{B}$ necessary to allow communication 
and to generate ${K}_{S}$.

We now suppose that Bob seeks to send a secret plaintext message $\varphi
^{M}$ to Alice via the public key ${K}_{P}$. On orderly measuring 
the particles $\mathcal{P}_{B}$ by using the public key ${K}_{P}$, 
Bob obtains the string $K_{B}=\{|k_{B}^{1}\rangle ,|k_{B}^{2}\rangle ,\cdots ,
|k_{B}^{n}\rangle \}$, where $|k_{B}^{i}\rangle = k_p^i p_B^i \in 
\{|0_{\hat{n}_{i}}\rangle , |1_{\hat{n}_{i}}\rangle \}$ are eigenstates of 
$\sigma_{\hat{n}_{i}}$. The message $\varphi ^{M}$ is characterized by a string 
of qubits $\varphi ^{M}=\{|\varphi _{P}^{1}\rangle,|\varphi _{P}^{2}\rangle ,\cdots ,
|\varphi _{P}^{n}\rangle \}$, where $|\varphi_{P}^{i}\rangle=\alpha_i|0\rangle+
\beta_i|1\rangle$ for $i\in \{1,2,\cdots, n\}$. Then Bob shall encrypt the message 
by applying a single qubit gate $G_{i}\in G=\{G_1, G_2, \cdots, G_n \}$ via
\begin{eqnarray}
|c^{i}\rangle=G_{i}|\varphi_{P}^{i}\rangle \label{eq:encryption}
\end{eqnarray}
where $G_{i}=H$ if $|k_{B}^{i}\rangle=|0_{\hat{n}_{i}}\rangle$  and 
$G_i=Z$ ($Z=\sigma_z$ is the Z-gate) in the other case if 
$|k_{B}^{i}\rangle=|1_{\hat{n}_{i}}\rangle$.
Thus the qubits $|c^{i}\rangle $ in the ciphertext ${C}$ are 
strongly dependent on the public key. We note that the general encryption
procedures, i.e. the general rule for choosing $H-$ and $Z-$gates are equally 
publicly announced.

The aim of the decryption algorithm is to decrypt the ciphertext ${C}$ and to recover
the plaintext $\varphi^M$ under the control of the private key. 
Since the private key $K_{S}$ is dynamical for our algorithm, Alice needs to obtain 
the private key $K_{S}$ prior to decrypting the ciphertext. 
Alice knows the public key ${K}_{P}$ as well as the secret parameters $\mathcal{B}$
and $\mathcal{U}$ and is thus enabled to calculate the private key ${K}_{S}$
by Eq.~(\ref{eq:secret}).
Then, Alice is required to measure the string of particles  $\mathcal{P}_{A}$ using the 
private key ${K}_{S}$ and obtains the secret string $K_{A}=\{|k_{A}^{1}\rangle ,
|k_{A}^{2}\rangle ,\cdots ,|k_{A}^{n}\rangle \}$ with $k_{A}^{i} =  k_{s}^{i} p_A^i$
for $i \in \{1, \cdots, n\}$.
Then Alice is in the position to evaluate Bob's measurement outcomes $K_{B}$ via $K_{A}$ because of the 
correlation of the measurement operators and the knowledge of the secret quantum channels $\mathcal{B}$
and the set of rotation operators $\mathcal{U}$. Say for example $|b_j\rangle=|\phi ^{+}\rangle$ 
may be the $j^{th}$ quantum channel and $k_p^j=\sigma_x$ the $j^{th}$ measurement operator of the 
public key for a particular $j \in \{1, 2, \cdots, n\}$. Then from Eq. (1) we learn that 
$|\phi ^{+}\rangle  = |+,0\rangle + |-,1\rangle$ in the basis of eigenstates of  $k_p^j=\sigma_x$ for Bob's 
particle, where the first and second entry of the MES refer to Bob's and Alice's particle, respectively.
As a consequence the possible outcomes for Bob's measurement via $k_p^j=\sigma_x$ could be either 
$k_B^j= |+\rangle$ or  $k_B^j= |-\rangle$. From table \ref{tab:correl} Alice knows the correlated measurement
operator $k_{s}^{j}=\sigma_z$. If her measurement $k_{A}^{j} =  \sigma_z p_A^j$ delivers 
$|0\rangle$, e.g., Bob's measurement must have resulted in $|+\rangle$, otherwise in $|-\rangle$.
Consequently, Alice can obtain Bobs set of qubit gates $G$ and thus decrypt the plaintext via
\begin{equation}
|\varphi_{P}^{i}\rangle= G_{i}^{\dagger} |c^i\rangle \label{eq:decryption}
\end{equation}
where $G_{i}^{\dagger} \in \{G_1^{\dagger}, G_2^{\dagger}, \cdots, G_n^{\dagger}\}$ 
are the adjoint operators of  $G_{i}$ as employed in Eq. (\ref{eq:encryption})
for $i \in \{1, \cdots, n\}$. We note that the $H-$ and $Z-$gates may be easily inverted.

\begin{figure}[htp]
\begin{center}
\includegraphics[width=160mm,height=74mm]{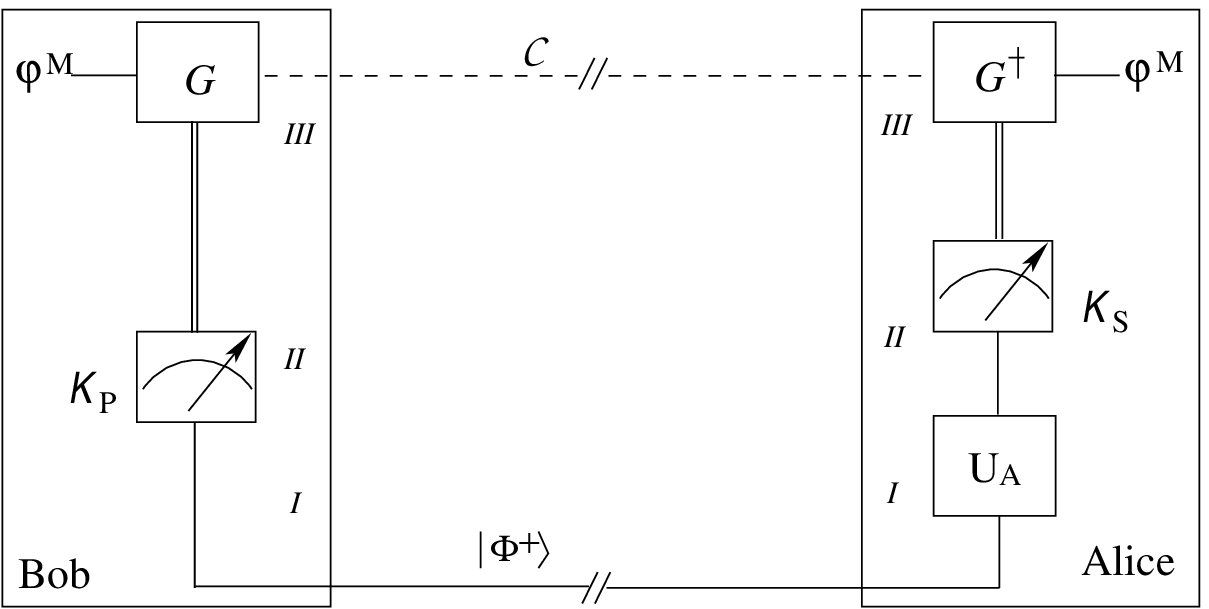}
\end{center}
\vspace{1.5cm}
\caption{Diagram of the quantum public key algorithm. The procedures of the encryption and decryption are divided into
three phases. In phase I a MES $|\Phi ^{+}\rangle$ is established between Alice and Bob, and then, Alice applies a
random unitary operation from $U_{A}$ on her particle of the entangled pair, which creates one of the eight
quantum channels. In phase II Bob and Alice perform measurements on their particles using the public and secret keys
$K_P$ and $K_S$,
respectively. For encryption and decryption in phase III, the key-dependent quantum logic gates in $G$ and $G^{\dagger}$ 
are applied on the plaintext $\varphi^{M}$ and the ciphertext $C$ by Bob and Alice, respectively.} 
\label{fig:algorithm}
\end{figure}

The above algorithm is illustrated in Fig.~\ref{fig:algorithm}, which includes the encryption and decryption
processes. The aim of the phases I and II is to establish the quantum channels between the particles of
the communicators and to carry out public and secret key-dependent measurements  on  
Bob's and Alice's particles, respectively. The resulting states of the measurements determine the set of quantum 
logic gates $G$ for the encoding and decoding procedures in phase III. 
Qubits as well as classic bits  may be encoded and decoded cryptographically this way. 
We emphasize further that the plaintext $\varphi ^{M}$ may be blocked for practical applications, 
when the number of bits of the plaintext exceeds that of the public key ${K}_{P}$. In this
situation Bob is required to divide the plaintext into $L$ blocks with length each of the public key $n$. 
Then he encrypts each qubit of the $i^{th}$ block for $i\in \{1,2,\cdots, L\}$ following the encryption 
procedure presented in Eq.~(\ref{eq:encryption}). If the whole plaintext or its last block are shorter 
than the public key, one should add some identity symbols, e.g. $|0\rangle $'s, alike
in classic communication, prior to encrypting this part of the plaintext. Similarly for the decryption, 
Alice repeats the decryption operation presented in Eq.~(\ref{eq:decryption}) for each block until all 
blocks have been decrypted.

We move on with the analysis of the security of the proposed
algorithm. In modern cryptography, the main characteristic is that
the encryption and decryption algorithms are public, while the
private key required for the actual decryption is secret. Thus 
the secrecy of the PKC depends completely on the secrecy of the 
private keys. As a consequence, an unconditionally secure algorithm 
requires it to be impossible for any attacker to obtain the private 
key neither directly nor through the public key, the cipher or any 
other insecurity of the algorithm. An attacker Eve may be an eavesdropper 
or a  tamper trying to modify the private key and shall not be assumed
here to be limited in resources in any way.

The first considered strategy of an attacker shall be to obtain or change the private 
key through the public key. Since ${K}_P$ is public, the attacker is obviously able 
to acquire it. 
The quantum channels, however, necessary to obtain ${K}_S$ via ${K}_P$ are nonorthogonal,
e.g. satisfy $|\langle \phi ^{+}|\Phi ^{+}\rangle |^{2}\neq 0$, which guarantees
that any attempt to intervene the quantum channel by an
eavesdropper Eve can be detected because of the noncloning
theorem \cite{wooters82}. Thus, the attacker, Eve, can not be part
of the quantum channel without disturbing it.
Moreover, according to table \ref{tab:correl}, there is  a probability of
$1/8$ for Eve of obtaining a single correct quantum channel. Thus, for an $n$ bit message
and the associated quantum channels, the probability for Eve of intercepting without being 
detected is $(1/8)^{n}$. This number becomes increasingly small for longer messages but more importantly
Eve may not know it even if she has found the correct quantum channels by accident. Those situations have 
been proven unconditionally secure (see first entry in \cite{bennett92} and references therein).  
In addition, due to the random variables $\theta_i$, there is no correlation between the public key 
and the private key. Thus without the knowledge of either $\mathcal{B}$ or one of $\mathcal{M}_{P}$ or 
$\mathcal{U}$, no information about ${K}_{S}$ is obtainable via ${K}_{P}$. 

Furthermore we consider the strategy, in which the attacker seeks to obtain the plaintext directly 
through the ciphertext. Since the ciphertext is created by the set of gates $G$ which is controlled
by $K_B$, this is required prior to finding the ciphertext. However, except for Bob and Alice, it is 
impossible for anybody to obtain the correct $K_{B}$ because after Bob's measurement on $\mathcal{P}_{B}$ 
using the public key there are two possible cases for each qubit. It is even impossible to acquire the 
correct ciphertext for any attacker, because the ciphertext consists of two states 
$\{Z|\psi^i_P\rangle, H|\psi^i_P\rangle\}$, which obey the property 
$|\langle \psi^i_P|Z^{\dagger}H|\psi^i_P\rangle|^2=\frac{1}{2}
\left[1+(\alpha^*_i\beta_i-\beta^*_i\alpha_i)\right]^2$. If $\alpha_i$ and $\beta_i$ are chosen to be real numbers,
then $|\langle \psi^i_P|Z^+H|\psi^i_P\rangle|^2=1/2$, which means these states are nonorthogonal. Thus the ciphertext may
not be identified like in the B92 protocol \cite{bennett84}. Accordingly, any qubit in the ciphertext is
unknown to the attacker, i.e. by the no-cloning theorem, the attacker can not copy or know it.

Unlike the classic PKC, whose security depends on the computational complexity assumption, the proposed algorithm does
not require such an assumption. It is implemented completely by the natural laws of quantum mechanics, i.e. does not involve
any intrinsic drawbacks. We add finally that usually the blocking treatment
decreases the security of the algorithm in the classic cryptography, because this treatment leaks some useful
information, such as the periodical or pseudo-periodical characteristics from the obtained ciphertext (consisting of
classic bits), to the attacker. However, the blocking treatment in the proposed algorithm does not disclose any
effective information, because no attacker is in the position to obtain  the correct ciphertext as mentioned above. 

In conclusion, an available and secure public key algorithm has been proposed.
The proposed algorithm encrypts the message using a public key and decrypts the
ciphertext using a private key. The public key may be publicly announced and
the private key is kept secret. Physically, the algorithm is implemented by
using correlations on the measurement axis of particles of a MES. The use of
the public key leads to a high availability, but it does not influence the
unconditional security of the proposed algorithm. The availability and the
unconditional security have been effectively united in the proposed algorithm.

This work is supported by an Alexander von Humboldt stipend for GZ (grant number 
IV CHN 1069575 STP) and the German Science Foundation (Nachwuchsgruppe within SFB 276).

\newpage

\begin{thebibliography}{}

\bibitem[Wiesner(1983]{wiesner83}
S. Wiesner, Sigact News. {\bf 15}, 78, (1983); C.H. Bennett {\em
et al.},
Advances in Cryptology: Proceedings of Crypto 82, 1982, edited by
D. Chaum, R.L. Rivest, and A.T. Sherman (Plenum Press, New York,
1982), p. 267.

\bibitem[Bennet(1984)]{bennett84}
C. H. Bennett, and G. Brassard, Advances in Cryptology:
Proceedings of Crypto'84, August 1984, Springer-Verlag, 475
(1984); C. H. Bennett, Phys. Rev. Lett., \textbf{68}, 3121,
(1992).

\bibitem[Ekert(1991)]{ekert91}
A. K. Ekert, Phys. Rev. Lett. \textbf{67}, 661, (1991).

\bibitem[Lo(1999)]{lo99}
H.-K. Lo and H.F.Chau, Science, {\bf 283}, 2050 (1999); P. W.
Shor, and J. Preskill, Phys. Rev. Lett. {\bf 85}, 441 (2000);
D. S. Naik \emph{et al.},
\emph{ibid} {\bf 84}, 4733 (2000).

\bibitem[Bennett(1992)]{bennett92}
C. H. Bennett \emph{et al.}, J. Crypto. {\bf 5}, 3 (1992); W. T.
Buttler \emph{et al.},
Phys. Rev. Lett. \textbf{84}, 5652 (2000).

\bibitem[Hillery(1999)]{hillery99}
M. Hillery, V. Buzek and A. Berthiaume, Phys. Rev. A {\bf 59},1829
(1999); R. Cleve, D. Gottesman, H. -K. Lo, Phys. Rev. Lett. {\bf
83}, 648 (1999).

\bibitem[Zeng(2000)]{zeng00}
G. Zeng, and W. Zhang, Phys. Rev A {\bf 61}, 032303 (2000).

\bibitem[Kent(1999)]{kent99}
A. Kent, Phys. Rev. Lett. {\bf 83}, 1447 (1999). H. P. Yuen,
quant-ph/0109055 and quant-ph/0106001; H. Buhrman, R. Cleve, J.
Watrous, and R. D. Wolf, Phys. Rev. Lett. {\bf 87}, 167902 (2001).

\bibitem[Terhal(2001)]{terhal01}
B. M. Terhal, D. P. DiVincenzo, and D. W Leung, Phys. Rev. Lett.
{\bf 86}, 5807 (2001); D. P. DiVincenzo, D. W Leung and B. M.
Terhal, arXiv: quant-ph/0103098.

\bibitem[Schumacher(1998)]{schumacher98}
B. Schumacher, Phys. Rev. Lett., {\bf 80}, 5695 (1998).

\bibitem[Schneier(1994)]{schneier94}
B. Schneier, \emph{Applied Cryptography: protocols, algorithms,
and source code in C} (John Wiley \& Sons, Inc., 1994)

\bibitem[Diffie(1976)]{diffie76} W. Diffie and M. E. Helman, IEEE Trans. Inf.
Theory {\bf 22}, 644 (1976);
R. L. Rivest, A. Shamir, and L.Adelman, Comm. ACM {\bf 21}, 120
(1978).

\bibitem[Chuang(1995)]{chuang95}
I. L. Chuang, R. Laflamme, P. W. Shor, and W. H. Zurek, Science
{\bf 270}, 1633 (1995).

\bibitem[Shor(1994)]{shor94} P. W. Shor, in Proceedings of the
35th Annual Symposium on FoCS (IEEE Press, Los Alamos, CA, 1994),
p.116.

\bibitem[Cirac(1995)]{cirac95}
J. I. Cirac and P. Zoller, Phys. Rev. Lett. {\bf 74}, 4091 (1995);
A. Barenco \emph{et al.},
\emph{ibid} {\bf 74}, 4083 (1995).

\bibitem[Wooters(1982)]{wooters82}
W. K. Wooters and W. H. Zurek, Nature
{\bf 299}, 802 (1982).

\end{thebibliography}
\end{document}